\documentstyle[12pt]{article}

\newcommand {\e} {\mbox{\rm e}}

\newcounter{eq}
\newcounter{sc}

\def\overleftrightarrow#1{\vbox{\ialign{##\crcr
 $\leftrightarrow$\crcr\noalign{\kern-1pt\nointerlineskip}
 $\hfil\displaystyle{#1}\hfil$\crcr}}}

\newlength{\minitwocolumn}
\begin{document}

\begin{flushright}
DPUR/TH/47\\
February, 2016\\
\end{flushright}
\vspace{20pt}

\pagestyle{empty}
\baselineskip15pt

\begin{center}
{\large\bf Induced Gravity and Topological Quantum Field Theory
\vskip 1mm }

\vspace{20mm}
Ichiro Oda \footnote{E-mail address:\ ioda@phys.u-ryukyu.ac.jp}

\vspace{5mm}
           Department of Physics, Faculty of Science, University of the 
           Ryukyus,\\
           Nishihara, Okinawa 903-0213, Japan.\\

\end{center}


\vspace{5mm}
\begin{abstract}
We construct an induced gravity (pregeometry) where both the Newton constant and the cosmological constant
appear as integration constants in solving field equations. By adding the kinetic terms of ghosts and 
antighosts, an action of the induced gravity is transformed to a topological field theory.
Moreover, by solving field equations of the topological field theory in the FRW universe, we find an inflation
solution. The present study might shed some light on a close relationship between the induced gravity and
the topological quantum field theory.     
\end{abstract}

\newpage
\pagestyle{plain}
\pagenumbering{arabic}


\rm
\section{Introduction}

Induced gravity, or pregeometry, has a long history whose origin traces back to a very
short paper by Sakharov in 1968 \cite{Sakharov}. The basic idea is that gravity is
not fundamental but might be induced by quantum fluctuations of matter fields. In particular,
the Newton constant could be induced at the one-loop level although it is vanishing
at the tree level. 

In the framework of the induced gravity, the existence of the Lorentzian manifold
is assumed $\it{a \ priori}$, but the dynamics of the geometry is not assumed and 
determined by radiative corrections of matter. Thus, the geometry is regarded as
a classical background and is not quantized unlike matter fields. Phrased in an equation, 
integrating over matter fluctuations at the one-loop level turns out to lead to an effective action
\begin{eqnarray}
S_{IG} = \frac{1}{16 \pi G} \int d^4 x  \sqrt{-g} \left( - 2 \Lambda + R 
+ \cdots \right),
\label{IG action}
\end{eqnarray}
where dots denote the higher-derivative terms such as $R^2$. Even if there are some difficulties and
subtleties in the induced gravity, the basic idea is extremely of interest in the sense
that quantum gravity is now believed to be somehow an emergent phenomenon like the induced 
gravity \cite{Seiberg}.
   
There are a number of versions of the induced gravity thus far \cite{Akama}-\cite{Frolov}, \footnote{See
a recent review on induced gravity \cite{Visser}.}
but the folklore of the induced gravity is to derive general relativity or its generalizations 
via quantum corrections of matter.
However,  there might be a logical possibility such that without relying on quantum fluctuations
gravity could be generated  at the classical level by solving field equations of a new type 
of induced gravity. For instance, in the pregeometrical approach of string field theory \cite{Hata, Horowitz}, 
starting with only a purely cubic action, the dynamics of the geometry or the kinetic term involving 
a background metric, is indeed induced by solving field equations and picking up a classical solution 
where quantum corrections play no role. 
One of motivations in this article is to construct such an induced gravity within the framework of a local field theory.     
  
In a recent work \cite{Oda0}, we have constructed a new induced gravity where the Newton constant is derived
by fixing a part of diffeomorphisms by which the full diffeomorphisms are broken down to the transverse 
diffeomorphisms (TDiff). However, it was difficult to derive the cosmological constant at the same time
since there remain no available diffeomorphisms for introducing a gauge condition corresponding to
the cosmological constant without violating the general covariance. In this paper, we will resolve this
issue and derive the Einstein equations with the cosmological constant by beginning with a new induced
gravity action.

It turns out that adding a sector of ghosts, we can transform the new induced gravity to 
a topological quantum field theory. Its moduli space is composed by constraints such that
the curvature density and the volume element are determined in terms of a divergence of vector densities.
One of these constraints is reminiscent of $\it{unimodular \ gravity}$ where the determinant of the metric
tensor is taken to be $-1$, that is, $\det g_{\mu\nu} = -1$ \cite{Einstein}-\cite{Saltas}.

Moreover, beginning with the topological theory, we derive the whole set of field equations and
find an inflation solution in the framework of the Friedmann-Robertson-Walker (FRW) universe with spacially 
flat metric.

The structure of this article is the following: In Section 2, we present a simple model of the induced gravity
and adding the ghost sector we construct a topological field theory. 
In Section 3, we examine the cosmological implications by finding the classical solution to the field
equations stemming from the topological model.
We conclude in Section 4.

\section{New model of induced gravity}

Let us start by presenting an action of our new induced gravity: 
\footnote{We follow notation and conventions by Misner et al.'s textbook \cite{MTW}, 
for instance, the flat Minkowski metric $\eta_{\mu\nu} = diag(-, +, +, +)$, the Riemann curvature tensor 
$R^\mu \ _{\nu\alpha\beta} = \partial_\alpha \Gamma^\mu_{\nu\beta} - \partial_\beta \Gamma^\mu_{\nu\alpha} 
+ \Gamma^\mu_{\sigma\alpha} \Gamma^\sigma_{\nu\beta} - \Gamma^\mu_{\sigma\beta} \Gamma^\sigma_{\nu\alpha}$, 
and the Ricci tensor $R_{\mu\nu} = R^\alpha \ _{\mu\alpha\nu}$. The reduced Planck mass is defined as 
$M_p = \sqrt{\frac{c \hbar}{8 \pi G}} = 2.4 \times 10^{18} GeV$ where $G$ is the Newton constant.}
\begin{eqnarray}
S = \int d^4 x  \sqrt{-g} \left[  \gamma \left( R - \varepsilon_0 \nabla_\mu \tau^\mu \right)
+ \lambda \left( 1 - v_0  \nabla_\mu \omega^\mu \right) \right],
\label{Action 1}
\end{eqnarray}
where $\gamma(x)$ and $\lambda(x)$ are the Lagrange multiplier fields enforcing the constraints 
$R = \varepsilon_0 \nabla_\mu \tau^\mu$ and $1 = v_0  \nabla_\mu \omega^\mu$, respectively. And
$v_0$ and $\varepsilon_0$ are some constants, and $\tau^\mu$ and $\omega^\mu$ are vector fields. 
The first term in this action was introduced to make the curvature density constraint $\sqrt{-g} R = 1$
be invariant under not the transverse diffeomorphisms (TDiff) but the full diffeomorphisms \cite{Oda0}.
The second term was also introduced to keep the unimodular constraint $\sqrt{-g} = 1$
be invariant under diffeomorphisms \cite{Henneaux}.

Now it is easy to show that the action (\ref{Action 1}) induces the Einstein-Hilbert action
with the cosmological constant by solving its field equations as follows: Variation with respect to
the vector fields reads
\begin{eqnarray}
\frac{\delta S}{\delta \tau^\mu} &=& \sqrt{-g} \varepsilon_0 \nabla_\mu \gamma = 0, \nonumber\\
\frac{\delta S}{\delta \omega^\mu} &=& \sqrt{-g} v_0 \nabla_\mu \lambda = 0,
\label{gamma-omega1}
\end{eqnarray}
from which the classical solution is given by
\begin{eqnarray}
 \gamma(x) = \bar \gamma, \quad \lambda(x) = \bar \lambda,
\label{gamma-omega2}
\end{eqnarray}
where $\bar \gamma$ and $\bar \lambda$ are certain constants. 
Then, provided that we set
\begin{eqnarray}
\bar \gamma = \frac{1}{16 \pi G}, \quad \bar \lambda = - \frac{2 \Lambda}{16 \pi G},
\label{gamma-omega3}
\end{eqnarray}
we obtain
\begin{eqnarray}
S = \frac{1}{16 \pi G} \int d^4 x  \sqrt{-g} \left( R - 2 \Lambda \right),
\label{EH Action}
\end{eqnarray}
which is nothing but the Einstein-Hilbert action with the cosmological constant $\Lambda$.
In this way, we can arrive at the action (\ref{EH Action}) of general relativity by
starting with the action (\ref{Action 1}) of the induced gravity simply by solving
field equations at the classical level. 

A peculiar feature of this new induced gravity is that the Newton constant ($G$)
and the cosmological constant ($\Lambda$) appear as integration constants which are 
not related to any parameters in the original induced gravity action. Let us recall 
that in unimodular gravity \cite{Einstein}-\cite{Saltas}, the similar phenomenon
emerges for the cosmological constant and is expected to play an important role in
understanding the well-known cosmological constant problem.

We are now ready to show that the induced gravity action (\ref{Action 1}) can be deformed into 
a topological quantum field theory by adding ghosts and antighosts as shown shortly.

First of all, let us consider two types of the BRST transformations. One type of the BRST
transformation $\delta_B$ is related to the first term in the action (\ref{Action 1}) and is given by
\begin{eqnarray}
\delta_B \gamma = \delta_B c = 0, \quad \delta_B \tau^\mu = \frac{1}{\varepsilon_0} \nabla^\mu c,
\quad \delta_B b = \gamma,
\label{BRST1}
\end{eqnarray}
where $c$ and $b$ are respectively a ghost with the ghost number $+1$ and an antighost with 
the ghost number $-1$. The Lagrange mutiplier field $\gamma$ is identified with the Nakanishi-Lautrup 
auxiliary field in this BRST transformation. The other BRST transformation $\hat{\delta}_B$
operates on the second term in the action (\ref{Action 1}) and takes the form   
\begin{eqnarray}
\hat{\delta}_B \lambda = \hat{\delta}_B \hat{c} = 0, \quad \hat{\delta}_B \omega^\mu 
= \frac{1}{v_0} \nabla^\mu \hat{c}, \quad \hat{\delta}_B \hat{b} = \lambda,
\label{BRST2}
\end{eqnarray}
where $\hat{c}$ and $\hat{b}$ are respectively a ghost with the ghost number $+1$ and an antighost with 
the ghost number $-1$. The Lagrange mutiplier field $\lambda$ corresponds to the Nakanishi-Lautrup 
auxiliary field as well. Note that the two types of the BRST transformations are nilpotent, $\delta_B^2 
= \hat{\delta}_B^2 = 0$ and anticommute to each other, $\{ \delta_B, \hat{\delta}_B \} = 0$, 
so we can define the physical state, $|phys \rangle$ by the physical state conditions, $Q_B |phys \rangle 
= \hat{Q}_B |phys \rangle = 0$ where $Q_B$ and $\hat{Q}_B$ are the BRST charges \cite{Kugo}.   

Next, we add the kinetic terms for the ghosts to the action (\ref{Action 1}). As a result, the total
action $S_T$ is given by
\begin{eqnarray}
S_T &=& S + \int d^4 x  \sqrt{-g} \left( - \nabla_\mu b \nabla^\mu c 
- \nabla_\mu \hat{b} \nabla^\mu \hat{c}  \right)   \nonumber\\
&=& \int d^4 x  \sqrt{-g} \left[  \gamma \left( R - \varepsilon_0 \nabla_\mu \tau^\mu \right)
- \nabla_\mu b \nabla^\mu c + \lambda \left( 1 - v_0  \nabla_\mu \omega^\mu \right) 
- \nabla_\mu \hat{b} \nabla^\mu \hat{c} \right]    \nonumber\\
&=& \int d^4 x  \sqrt{-g} \left\{ \delta_B \left[  b \left( R - \varepsilon_0 \nabla_\mu \tau^\mu \right)\right]
+ \hat{\delta}_B \left[ \hat{b} \left( 1 - v_0  \nabla_\mu \omega^\mu \right) \right] \right\}.
\label{Total Action}
\end{eqnarray}
The last expression clearly implies that $S_T$ is an action of the topological quantum field theory
where its moduli space is defined by two equations $R = \varepsilon_0 \nabla_\mu \tau^\mu$ and
$1 = v_0 \nabla_\mu \omega^\mu$.

About twenty five years ago, we have constructed a model of topological pregeomery where a classical action 
has been taken to be trivially zero and quantum fluctuations of matter fields have generated the Einstein-Hilbert 
action with the cosmological constant at the one-loop level via the cutoff which violates the topological 
symmetry  \cite{Oda1, Oda2}. 

On the other hand, in this paper, we have started with a non-trivial classical action (\ref{Action 1}) 
of the induced gravity, which reduces to the Einstein-Hilbert action with the cosmological constant at the
classical level, and then added the kinetic term of the ghosts, thereby transforming the induced gravity
to a topological field theory. The above two models of the induced gravity are obviously different, but
the close relationship between the induced gravity and the topological field theory seems to suggest that the
induced gravity and a topological quantum field theory might be tantalizingly equivalent.

\section{Cosmological solutions}

In this section, we work with the action (\ref{Total Action}) of a topological field theory to
derive the inflation universe as a classical solution in the framework of the Friedmann-Robertson-Walker 
(FRW) universe with spacially flat metric. This derivation not only shows indirectly that the topological action 
(\ref{Total Action}) is equivalent to that of general relativity but also suggests that it could be applied to
cosmology, in particular, the inflation universe.

The Einstein equations, which stem from variation with respect to the metric tensor $g^{\mu\nu}$, read
\begin{eqnarray}
&{}& 2 \gamma G_{\mu\nu} - 2 \left( \nabla_\mu \nabla_\nu \gamma - g_{\mu\nu} \nabla^2 \gamma \right)
+ 2 \left[ \varepsilon_0 \tau_{(\mu} \nabla_{\nu)} \gamma - \nabla_{(\mu} b \nabla_{\nu)} c
+ v_0 \omega_{(\mu} \nabla_{\nu)} \lambda - \nabla_{(\mu} \hat{b} \nabla_{\nu)} \hat{c} \right]
\nonumber\\
&{}& - g_{\mu\nu} \left[ \lambda + \varepsilon_0 \tau^\rho \nabla_\rho \gamma - \nabla_\rho b \nabla^\rho c
+ v_0 \omega^\rho \nabla_\rho \lambda - \nabla_\rho \hat{b} \nabla^\rho \hat{c} \right]
= 0,
\label{Einstein-eq 1}
\end{eqnarray}
where $G_{\mu\nu} = R_{\mu\nu} - \frac{1}{2} g_{\mu\nu} R$ is the Einstein tensor and the round bracket indicates 
the symmetrization of indices of weight $\frac{1}{2}$ such as $A_{(\mu} B_{\nu)} = \frac{1}{2} \left( A_\mu B_\nu
+ A_\nu B_\mu \right)$. In deriving this equation, by performing the integration by parts, it is convenient to rewrite 
the action (\ref{Total Action}) as
\begin{eqnarray}
S_T = \int d^4 x  \sqrt{-g} \left[  \gamma R + \varepsilon_0 \tau^\mu \nabla_\mu \gamma
- \nabla_\mu b \nabla^\mu c + \lambda + v_0  \omega^\mu \nabla_\mu \lambda 
- \nabla_\mu \hat{b} \nabla^\mu \hat{c} \right].
\label{Total Action 2}
\end{eqnarray}

Taking variation of the vector fields $\tau^\mu, \omega^\mu$ gives rise to Eq. (\ref{gamma-omega1})
whose solution is given by Eq. (\ref{gamma-omega2}). The field equations for the Lagrange multiplier fields 
$\gamma$ read
\begin{eqnarray}
R = \varepsilon_0 \nabla_\mu \tau^\mu, 
\label{gamma1}
\end{eqnarray}
which can be rewritten as
\begin{eqnarray}
\sqrt{-g} R = \varepsilon_0 \partial_\mu \left( \sqrt{-g} \tau^\mu \right).
\label{gamma2}
\end{eqnarray}
Likewise, for the $\lambda$ variation, we have
\begin{eqnarray}
1 = v_0 \nabla_\mu \omega^\mu,
\label{lambda1}
\end{eqnarray}
which can be recast to the form
\begin{eqnarray}
\sqrt{-g} = v_0 \partial_\mu \left( \sqrt{-g} \omega^\mu \right).
\label{lambda2}
\end{eqnarray}
Eqs. (\ref{gamma2}) and (\ref{lambda2}) imply that the scalar curvature density ($\sqrt{-g} R$)
and the volume element ($\sqrt{-g}$) are determined by a divergence of the vector densities $\sqrt{-g} \tau^\mu$
and $\sqrt{-g} \omega^\mu$, respectively.

Finally, field equations for the ghosts and the antighosts satisfy the same form of equation
$\nabla_\mu \nabla^\mu \Phi = 0$ where $\Phi = \{ c, b, \hat{c}, \hat{b} \}$. Since the ghosts and the
antighosts carry non-zero ghost numbers, we simply take $c = b = \hat{c} = \hat{b} = 0$ as the classical
solution.

With the solution Eq. (\ref{gamma-omega2}) and the vanishing ghosts and antighosts, 
the Einstein equations (\ref{Einstein-eq 1}) takes the simpler form
\begin{eqnarray}
G_{\mu\nu} - \frac{\bar \lambda}{2 \bar \gamma} g_{\mu\nu} = 0,
\label{Einstein-eq 2}
\end{eqnarray}
which is equivalent to the standard Einstein equations with the cosmological constant $\Lambda 
= - \frac{\bar \lambda}{2 \bar \gamma}$.

At this stage, having the cosmological application of the model at hand in mind, let us 
work with the FRW metric with spacially flat metric ($k = 0$)
\begin{eqnarray}
d s^2 = g_{\mu\nu} d x^\mu d x^\nu = - d t^2 + a(t)^2 ( d x^2 + d y^2 + d z^2 ),
\label{FRW metric}
\end{eqnarray}
with $a(t)$ being the scale factor. Since the equations (\ref{Einstein-eq 2}) are usual
Einstein equations with the cosmological constant, the Hubble parameter $H = \frac{\dot{a}}{a}$ 
takes a constant value $H = \bar H$ where $\bar H$ is a constant. Then, the scale factor 
is of form $a(t) = a(0) \e^{\bar H t}$ so that we have the inflation universe.

The remaining field equations (\ref{gamma1}) and (\ref{lambda1}) must be solved. 
In order to solve these equations in an analytical way, we assume that
\begin{eqnarray}
\tau_0 &=& \tau_0(t), \quad \dot{\tau_0} = \tau_i = 0,   \nonumber\\
\omega_0 &=& \omega_0(t), \quad \dot{\omega_0} = \omega_i = 0,
\label{Ansatz}
\end{eqnarray}
where the overdot means the differentiation with respect to the time variable and
the index $i = 1, 2, 3$ denotes the space component. Then, Eq. (\ref{gamma1}) yields
the equation
\begin{eqnarray}
\bar H = - \frac{1}{4} \varepsilon_0 \tau_0,
\label{gamma sol}
\end{eqnarray}
and Eq. (\ref{lambda1}) is solved to be
\begin{eqnarray}
\bar H = - \frac{1}{3} \frac{1}{v_0 \omega_0}.
\label{lambda sol}
\end{eqnarray}
Hence, provided that we make use of Eq. (\ref{gamma-omega3}) and set $\varepsilon_0 = v_0 = 1$,
the time components $\tau_0, \omega_0$ of the vector fields are described in terms of the
cosmological constant $\Lambda$ 
\begin{eqnarray}
\tau_0 = - 4 \sqrt{\frac{\Lambda}{3}}, \quad \omega_0 = - \frac{1}{\sqrt{3 \Lambda}}.
\label{gamma-lambda sol}
\end{eqnarray}

\section{Conclusion}

In this article, motivated with unimodular gravity, we have constructed a new type of induced gravity.
In unimodular gravity, the cosmological constant emerges as an integration constant whereas
in our induced gravity, both the Newton constant and the cosmological constant appear as integration
constants. This physically intriguing feature is utilized to derive general relativity from
the induced gravity at the classical level. We think that our induced gravity is an analog
of the pregeometrical approach of string field theory in local field theory.

Furthermore, we have pursued our original idea that the induced gravity is closely
connected with the topological field theory and shown that indeed the induced gravity
considered in this paper can be deformed into a topological field theory by adding
the kinetic terms of ghosts.  

We have also found the inflation solution by solving analytically the field equations 
of the topological field theory. Of course, adding additional matter fields in our model, it is
possible to derive various types of cosmological solutions as in general relativity.
However, the incorporation of matter fields is against our philosophy since matter
fields break the topological symmetry.

In this article, we have treated with the metric tensor field as a classical background
field along the spirit of the induced gravity, but it might be interesting to regard
the metric tensor as a quantized field as well and investigate its physical implications 
\footnote{We have already made a model of topological gravity in four dimensions 
\cite{Oda3, Oda4}.}

\begin{flushleft}
{\bf Acknowledgements}
\end{flushleft}
This work is supported in part by the Grant-in-Aid for Scientific 
Research (C) No. 25400262 from the Japan Ministry of Education, Culture, 
Sports, Science and Technology.



\begin{thebibliography}{99}

\bibitem{Sakharov}
A. D. Sakharov, {Sov. Phys. Dokl. {\bf 12} (1968) 1040.} 	

\bibitem{Seiberg}
N. Seiberg, {arXiv:hep-th/0601234.}
 
\bibitem{Akama}
K. Akama, Y. Chikashige, T. Matsuki and H. Terazawa, {Prog. Theor. Phys. {\bf 60} (1978) 868.}

\bibitem{Adler2}
S. L. Adler, {Phys. Lett. {\bf B 95} (1980) 241.}

\bibitem{Zee}
A. Zee, {Phys. Rev. {\bf D 23} (1981) 858.}

\bibitem{Amati}
D. Amati and G. Veneziano, {Nucl. Phys. {\bf B 204} (1982) 451.}

\bibitem{Adler1}
S. L. Adler, {Rev. Mod. Phys. {\bf 54} (1982) 729.}
 	
\bibitem{Oda1}
K. Akama and I. Oda, {Phys. Lett. {\bf B 259} (1991) 431.}

\bibitem{Oda2}
K. Akama and I. Oda, {Nucl. Phys. {\bf B 397} (1993) 727.}

\bibitem{Frolov}
V. P. Frolov, D. V. Fursaev and A. I. Zelnikov, {Nucl. Phys. Proc. 
Suppl. {\bf 57} (1997) 192.}

\bibitem{Visser}
M. Visser, {Mod. Phys. Lett. {\bf A 17} (2002) 977.}

\bibitem{Hata}
H. Hata, K. Itoh, T. Kugo, H. Kunitomo and K. Ogawa, {Phys. Lett. {\bf B 175} (1986) 138.}

\bibitem{Horowitz}
G. Horowitz, J. Lykken, R. Rohm and A. Strominger, {Phys. Rev. Lett. {\bf 57} (1986) 283.}

\bibitem{Oda0}
I. Oda, {arXiv:1602.00851 [gr-qc].}
 
\bibitem{Einstein}
A. Einstein, {in {"The Principle of Relativity", by A. Einstein et al., Dover
Publications, New York, 1952.}}

\bibitem{Bij}
J. van der Bij, H. van Dam and Y. J. Ng, {Physica {\bf 116A} (1982) 307.}

\bibitem{Buchmuller1}
W. Buchmuller and N. Dragon, {Phys. Lett. {\bf B 207} (1988) 292.}

\bibitem{Henneaux}
M. Henneaux and C. Teitelboim, {Phys. Lett. {\bf B 222} (1989) 195.}

\bibitem{Buchmuller2}
W. Buchmuller and N. Dragon, {Phys. Lett. {\bf B 223} (1989) 313.}

\bibitem{Unruh}
W. G. Unruh, {Phys. Rev. {\bf D 40} (1989) 1048.}

\bibitem{Ng}
Y. J. Ng and H. van Dam, {J. Math. Phys. {\bf 32} (1991) 1337.}

\bibitem{Faedo1}
E. Alvarez and A. F. Faedo, {Phys. Rev. {\bf D 76} (2007) 064013.}

\bibitem{Faedo2}
E. Alvarez, A. F. Faedo and J. J. Lopez-Villarejo, {JHEP {\bf 0810} (2008) 023.}

\bibitem{Smolin}
L. Smolin, {Phys. Rev. {\bf D 80} (2009) 084003.}

\bibitem{Ellis}
G. F. R. Ellis, H. van Elst, J. Murugan and J. -P. Uzan, {Class. Quant. Grav. 
{\bf 28} (2011) 225007.}

\bibitem{Padilla}
A. Padilla and I. D. Saltas, {Eur. Phys. J. {\bf C 75} (2015) 561.}

\bibitem{Saltas}
I. D. Saltas, {Phys. Rev. {\bf D 90} (2014) 124052.}

\bibitem{MTW}
C. W. Misner, K. S. Thorne and J. A. Wheeler, {"Gravitation", W H Freeman and Co (Sd), 1973.}

\bibitem{Kugo}
T. Kugo and I. Ojima, {Prog. Theor. Phys. Suppl. {\bf 66} (1979) 1.}

\bibitem{Oda3}
I. Oda and A. Sugamoto, {Phys. Lett. {\bf B 266} (1991) 280.}

\bibitem{Oda4}
A. Nakamichi, A. Sugamoto and I. Oda, {Phys. Rev. {\bf D 44} (1991) 3835.}
 	

\end{thebibliography}
\end{document}